\documentclass[]{spie}  
\usepackage[dvips]{graphicx}

\title{Physics and Operation of the AMANDA-II High Energy Neutrino Telescope} 

\author{Steven W. Barwick, for the AMANDA Collaboration}

\authorinfo{Slightly updated contribution to SPIE in Kona, Hawaii, 2002\\S.W.B.: E-mail:barwick@cosmic.ps.uci.edu, Telephone: 1 949 824 2626\\ Address: University of California, Irvine, CA 92697}

\begin{document} 
  \maketitle 
	%
%
\begin{sloppypar}

\noindent
J.~Ahrens$^{11}$, 
X.~Bai$^{1}$, 
S.W.~Barwick$^{10}$, 
T.~Becka$^{11}$, 
K.-H.~Becker$^{2}$, 
E.~Bernardini$^{4}$,
D.~Bertrand$^{3}$, 
F.~Binon$^{3}$, 
A.~Biron$^{4}$, 
S.~B\"oser$^{4}$
% ...

\end{sloppypar}

\vspace*{0.2cm}

{\footnotesize
\noindent
   (1) Bartol Research Institute, University of Delaware, Newark, DE 19716, USA
   \newline
   (2) Fachbereich 8 Physik, BUGH Wuppertal, D-42097 Wuppertal, Germany
   \newline
   (3) Universit\'e Libre de Bruxelles, Science Faculty CP230, Boulevard du Triomphe, B-1050 Brussels, Belgium
   \newline
   (4) DESY-Zeuthen, D-15735 Zeuthen, Germany
   \newline
   (6) Dept. of Technology, Kalmar University, S-39182 Kalmar, Sweden
   \newline
   (7) Lawrence Berkeley National Laboratory, Berkeley, CA 94720, USA
   \newline
   (8) Dept. of Physics, Pennsylvania State University, University Park, PA 16802, USA
   \newline
   (9) Dept. of Physics, University of California, Berkeley, CA 94720, USA
   \newline
   (10) Dept. of Physics and Astronomy, University of California, Irvine, CA 92697, USA
   \newline
   (11) Institute of Physics, University of Mainz, Staudinger Weg 7, D-55099 Mainz, Germany
   \newline
   (12) University of Mons-Hainaut, 7000 Mons, Belgium
   \newline
   (13) Departamento de F\'{\i}sica, Universidad Sim\'on Bol\'{\i}var, Apdo. Postal 89000, Caracas, Venezuela
   \newline
   (14) Dept. of Physics, University of Wisconsin, Madison, WI 53706, USA
   \newline
   (15) Physics Dept., University of Wisconsin, River Falls, WI 54022, USA
   \newline
   (16) Division of High Energy Physics, Uppsala University, S-75121 Uppsala, Sweden
   \newline
   (17) Dept. of Physics, Stockholm University, SCFAB, SE-10691 Stockholm, Sweden
   \newline
   (18) Vrije Universiteit Brussel, Dienst ELEM, B-1050 Brussel, Belgium
   \newline
   (19) Present address: Santa Cruz Institute for Particle Physics, University of California, Santa Cruz, CA 95064, USA
   \newline
   (20) Present address: CERN, CH-1211, Gen\`eve 23, Switzerland
   \newline
}

\begin{abstract}
This paper briefly describes the principle of operation and science goals of the AMANDA high energy neutrino telescope located at the South Pole, Antarctica.  Results from an earlier phase of the telescope, called AMANDA-B10, demonstrate both reliable operation and the broad astrophysical reach of this device, which includes searches for a variety of sources of ultrahigh energy neutrinos: generic point sources, Gamma-Ray Bursts and diffuse sources.  The predicted sensitivity and angular resolution of the telescope were confirmed by studies of atmospheric muon and neutrino backgrounds. We also report on the status of the analysis from AMANDA-II, a larger version with far greater capabilities. At this stage of analysis,  details of the ice properties and other systematic uncertainties of the AMANDA-II telescope are under study, but we have made progress toward critical science objectives.  In particular, we present the first preliminary flux limits from AMANDA-II on the search for continuous emission from astrophysical point sources, and report on the search for correlated neutrino emission from Gamma Ray Bursts detected by BATSE before decommissioning in May 2000.  During the next two years, we expect to exploit the full potential of AMANDA-II with the installation of a new data acquisition system that records full waveforms from the in-ice optical sensors. 
\vspace{1pc}
\end{abstract}



\section{INTRODUCTION}
\label{sect:intro}  

Nature provides precious few information carriers from the deep recesses of space, and it is imperative to develop techniques to exploit them all. Throughout history, the photon messenger has made vital contributions to the understanding of the observable Universe. In this paper, we present results from a new generation of telescope designed to detect a very different kind of information carrier, high energy neutrinos (where $E_{\nu}>$ 1 TeV).  The search for astronomical sources of high energy neutrinos is one of the central missions of the Antarctic Muon and Neutrino Detector Array (AMANDA), located close to the geographic South Pole in Antarctica~\cite{Aman99}.  It is possible to identify sources (i.e, astronomy) with neutrinos because they are neutral and stable. The most powerful detection technique relies on the observation of a muon that is created by charged-current interactions between a neutrino and atomic nuclei in and around the instrumented volume of the detector.  At the energies of interest, the direction of the muon is highly correlated with the direction of the parent neutrino.  The correlation is usually characterized by the mean angle of deviation, $\theta_{\nu\mu}\sim 0.65^{o}/(E_{\nu})^{0.48}$, where $E_{\nu}$ is the neutrino energy in units of $10^{12}$ eV.  The correlation between the detected muon and neutrino direction is helped by the fact that the higher energy muons are more readily detected because they emit more Cherenkov photons and they travel further than lower energy ones.  Ideally, the muon angular resolution of the telescope should be comparable to or less than the angular correlation between muon and neutrino.  The muon energy can be estimated from the relative increase in Cherenkov emission mainly due to pair production and bremsstrahlung.

Ultrahigh energy (UHE) neutrinos with energies in the TeV range and
higher may be produced by a variety of sources.  Particle physics
exotica like WIMPs and topological defects are expected to produce
neutrinos in their annihilation or decay~\cite{uhe-nu-hep-theories},
and models of astrophysical phenomena such as gamma-ray bursts, active
galactic nuclei, supernovae and
microquasars~\cite{uhe-nu-astro-theories,Atoyan,mBlazar} also predict UHE neutrino
fluxes. Micro-black holes may be produced in the ice by collisions with extremely high energy neutrinos\cite{mBH}.

AMANDA is sensitive to UHE neutrinos produced by these sources and can
provide some of the most stringent tests to date of UHE neutrino
production models.  More generally, AMANDA and other similar neutrino
telescopes~\cite{other-experiments} open a heretofore unexplored
window on the universe in a region of the energy spectrum bounded
between roughly $10^{12}$~eV and $10^{20}$~eV.  In the somewhat
narrower energy range between roughly $10^{14}$~eV and $10^{19}$~eV,
photons are absorbed by intervening matter and starlight, and
cosmic-ray protons are insufficiently energetic to reach us without
experiencing unknown amounts of curvature in intervening magnetic
fields, leaving neutrinos as the only known particles that can serve
as astronomical messengers.  Neutrino telescopes are also sensitive to
supernova neutrino bursts at neutrino energies of roughly $10^7$~eV.

While the scientific potential of neutrino astronomy is broad, far reaching, and exciting, it is important to keep expectations realistic by assessing the scientific and technical capabilities of current and future neutrino telescopes. In the next section(s), we describe the detection technique, signatures, effective size, and backgrounds that define the capabilities of AMANDA-II. 

\section{THE AMANDA DETECTOR}
The essential characteristics of a neutrino telescope have been known for more than two decades and most important features were discussed and specified during a series of workshops devoted to developing the DUMAND concept.  In fact, more than four decades ago, Markov suggested that
the ocean would be a suitable site for constructing a large neutrino detector based on the detection of Cherenkov light. Halzen and Learned \cite{icehistory} introduced a 
twist on the general scheme by promoting polar ice as suitable medium. Until recently, workable implementations of these sensible ideas have been thwarted by unusual technical and logistical challenges associated
with the remote deployment of hardware in media that differ from ordinary purified water in several important details.

The AMANDA-B10 high energy neutrino detector consists of lattice of 302 optical
modules (OMs) on 10 strings.  Each OM is comprised of a photomultiplier
tube (PMT) with passive electronics housed in a glass pressure vessel.
The OMs are deployed within a cylindrical volume about 120~m in diameter
and 500~m in height at depths between roughly 1500 and 2000~m below
the surface of the South Pole ice cap.  In this region the optical
properties of the ice are well suited for reconstructing the Cherenkov
light pattern emitted by relativistic charged
particles~\cite{ice-properties}. This light is used to reconstruct
individual events.  An electrical cable provides high voltage to the
PMTs and transmits their signal pulses to the surface electronics.  A
light diffuser ball connected via fiber optic cable to a laser on the
surface is used for calibration purposes.  Copious down-going cosmic
ray muons are also used for calibration purposes.

\begin{figure}[htb]
\vspace{9pt}
\includegraphics[scale=0.4]{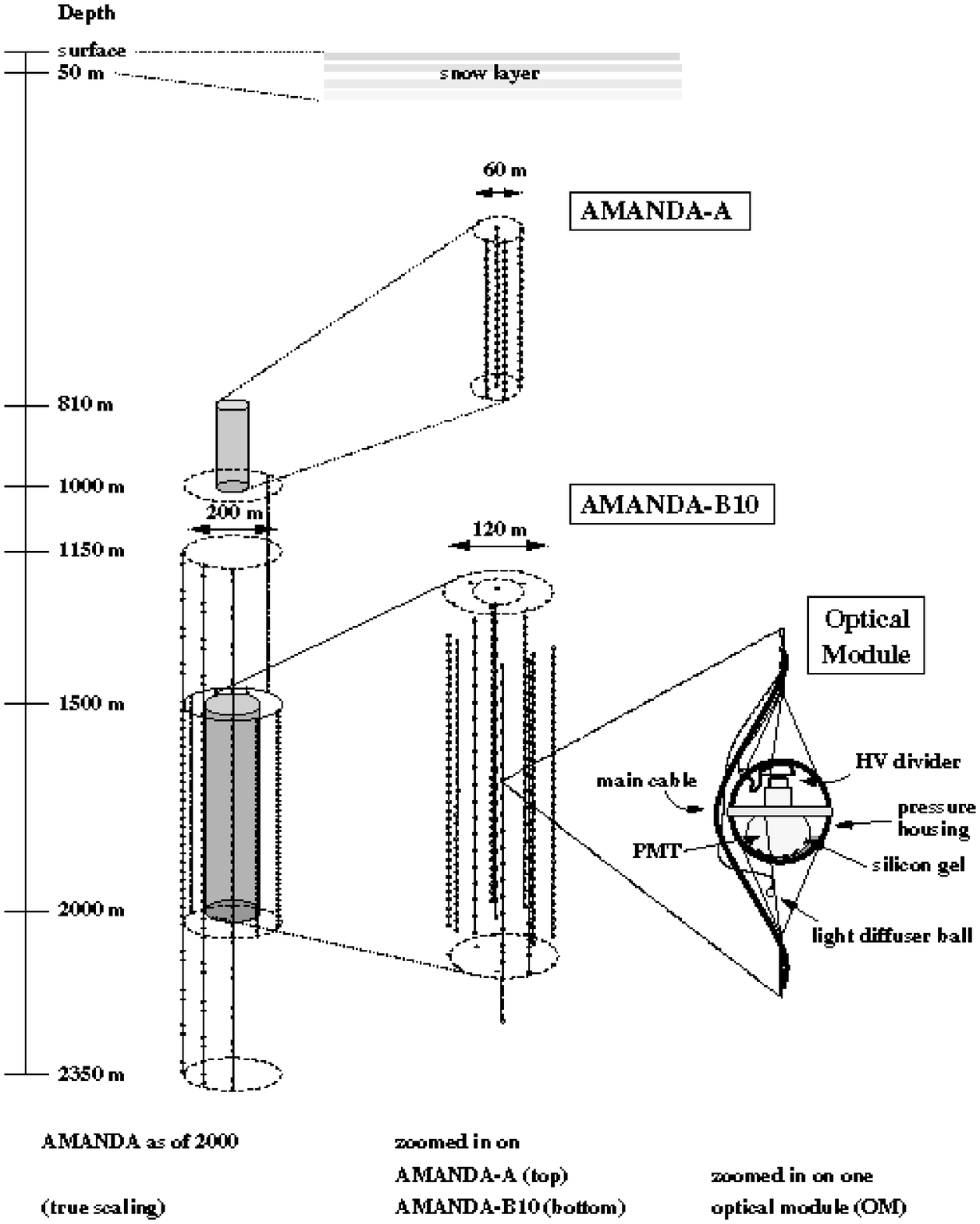}
\includegraphics[scale=0.4]{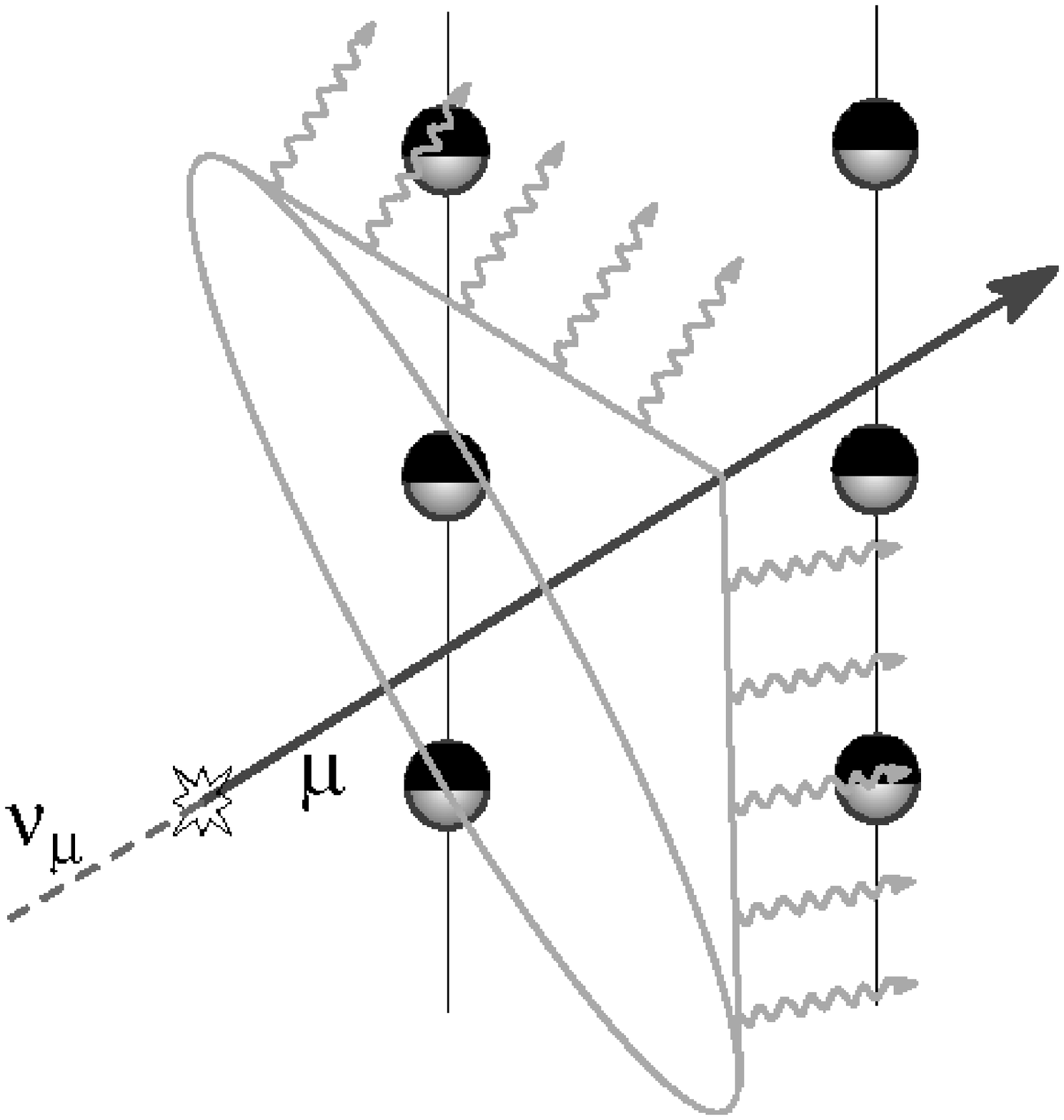}
\caption{{\bf Left: }AMANDA-B10 consists of 302 optical modules in a cylindrical volume 
120~m diameter and 500~m in height.  To build AMANDA-II, optical
modules were added to bring the count up to 667 OMs in a cylindrical
volume 200~m in diameter.}

\label{fig:amanda-detector}
\caption{{\bf Right: }Schematic of the detection method most commonly employed by the AMANDA-II high energy neutrino telescope.  A muon is produced by charged-current interactions initiated by muon neutrinos. The muon produces UV-blue Cherenkov light that is detected by the embedded optical sensors. Muon trajectory is reconstructed from the time of arrival and geometric information.}
\label{fig:cherenkov}

\end{figure}

In January 2000, AMANDA-B10 was enlarged to a total of 19 strings with
667 OMs to form AMANDA-II.  This new detector is 200~m in diameter and
approximately the same height and depth as AMANDA-B10.
Figure~\ref{fig:amanda-detector} shows a schematic diagram of AMANDA.
Planning, design, and construction of the new drill has begun on IceCube, a kilometer-scale device with 4800 OMs on 80 strings~\cite{albrechts-contribution}.

The optical sensors respond to the UV-blue dominated cherenkov radiation emitted by neutrino-induced muons (see Fig.~\ref{fig:cherenkov}) or neutrino-induced hadronic and electromagnetic cascades.  Large 
detector volumes are  required because the predicted flux of cosmic neutrinos and the known interaction probabilities at the energies of interest are relatively small. 
The detection probability, defined as the ratio between 
the range of the muon to the interaction mean free path of the neutrino,  is only 10$^{-6}$ for neutrinos with an energy of 1 TeV, which is small. Moreover, the rare signal events must be extracted from a large flux of atmospheric
muon background generated by interactions between cosmic rays and the nuclei in the atmosphere. To minimize this problem, AMANDA was deployed between 1.5 and 1.9 kilometers underneath the ice surface.  The required combination of large volume, large material overburden, and desire to minimize costs leaves 
few options other than to construct a detector within a 
remote, naturally occurring, transparent medium such as Antarctic ice or water (no excavated caves or mines are large enough). The formidable technical challenge of remote operation distinguishes high energy neutrino facilities from existing solar and accelerator-based neutrino detectors.  

Given these constraints, AMANDA technologies and system architecture were developed, and now proven by five years of operation, to be reliable, durable, and robust.  The technical capabilities of the embedded hardware were sufficient to accomplish to the primary science missions. 
AMANDA was shown to be a functioning neutrino {\textit detector} by virtue of
its ability to reconstruct upward-going muons induced by atmospheric
muon neutrinos~\cite{nature,b10-atmnu} and detailed comparison between experimental data and background simulations. It was shown to be a functioning {\textit telescope} by comparing the expected pointing accuracy and angular resolution with data obtained from air shower events that simultaneously trigger the SPASE air shower array and AMANDA.  

AMANDA-B10 data has also been used to set competitive limits on
WIMPs~\cite{b10-wimps}, monopoles~\cite{b10-monopoles}, extremely
energetic neutrinos~\cite{b10-ehe}, UHE $\nu_\mu$ point
sources~\cite{b10-point-source} and diffuse fluxes~\cite{b10-diffuse}.
The detector is also sensitive to
bursts of low energy neutrinos from
supernovae~\cite{b10-supernova}.  In the following sections, we highlight
some of the previously reported work, and discuss new analysis of AMANDA-II data.

\section{SEARCH FOR UHE $\nu_\mu$ FROM POINT SOURCES WITH AMANDA-II}

We have conducted a general search for continuous emission of muon
neutrinos from a spatially localized direction in the northern sky.
Backgrounds are reduced by requiring a statistically significant
enhancement in the number of reconstructed upward-going muons within a
small bin in solid angle. The bin size is determined from the predicted angular resolution for muon events.  Furthermore, the background for a
particular bin can be calculated from the data by averaging over the
data external to that bin in the same declination band.  In
contrast to other specialized analyses used by the AMANDA collaboration, this search is more tolerant of the presence of background, so the signal is optimized on $S/\sqrt{B}$,
where $S$ represents the signal and $B$ the background, rather than on
$S/B$, which emphasizes signal purity.

Data acquired by AMANDA-B10 in 1997 has been analyzed and the results
presented~\cite{b10-point-source}. That paper details several checks of the
simulation programs, such as using SPASE-AMANDA coincidence data to check the 
absolute pointing and angular resolution, and comparing background predictions with data from trigger to the final selection criteria. Systematic uncertainties in simulation input parameters were studied and included in the final results.

\begin{figure}[htb]
\centering
\includegraphics[scale=0.6]{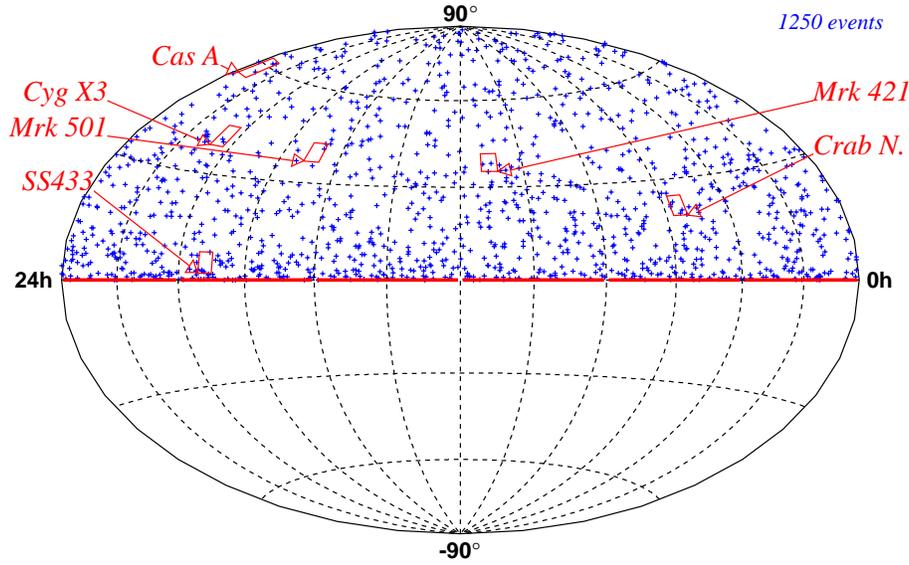}
\caption{Sky plot obtained from the AMANDA-II point source analysis. Horizontal coordinates are right ascension and vertical coordinates are declination.  Also shown are the sky coordinates and characteristic search bin for several potential neutrino sources.}
\label{fig:allsky}
\end{figure}

 With AMANDA-II data taken in
2000, we gain improved sensitivity over the entire visible sky, and most critically,  to events near the horizon since
the detector has double the number of PMTs and a larger lever arm in
the horizontal dimension.  This is illustrated by the 1250 events in Fig. ~\ref{fig:allsky}.  With the exception of the horizon, the event distribution is uniform in declination. In order to achieve
blindness in this analysis the right ascension of each event (i.e., its azimuthal angle) was scrambled (at the South Pole this effectively
scrambles the event time) before the analysis was finalized. The goal was to search for a statistically significant excess of events from a specific direction in the sky. The analysis divided the sky into non-overlapping $6^\circ \times 6^\circ$ angular bins at the horizon, and varied the azimuthal width of the bin at larger declinations to maintain approximately constant solid angle. The width of the bin is roughly a factor of 3 larger than the angular resolution of the detector. Four distinct maps were created to maximize the counts for sources that lie near the edge in a given map. The maps differ by shifting the center of each bin by half the width.  All sky bins are consistent with random fluctuation of the background events that remain in the sample. Using a customary source spectrum proportional to $E^{-2}$, preliminary flux limits for several sources are presented in Table~\ref{table:point_source_sensitivities}.

In addition to the preliminary limits extracted from data collected in 2000, Table~\ref{table:point_source_sensitivities} presents projected sensitivities of AMANDA-II using all the data currently on tape. The sensitivity is defined as the predicted average 90\% C.L. limit from an
ensemble of experiments with no signal, and is calculated using background levels predicted from off-source data.

\begin{figure}[htb]
\vspace{9pt}
\includegraphics[scale=0.6]{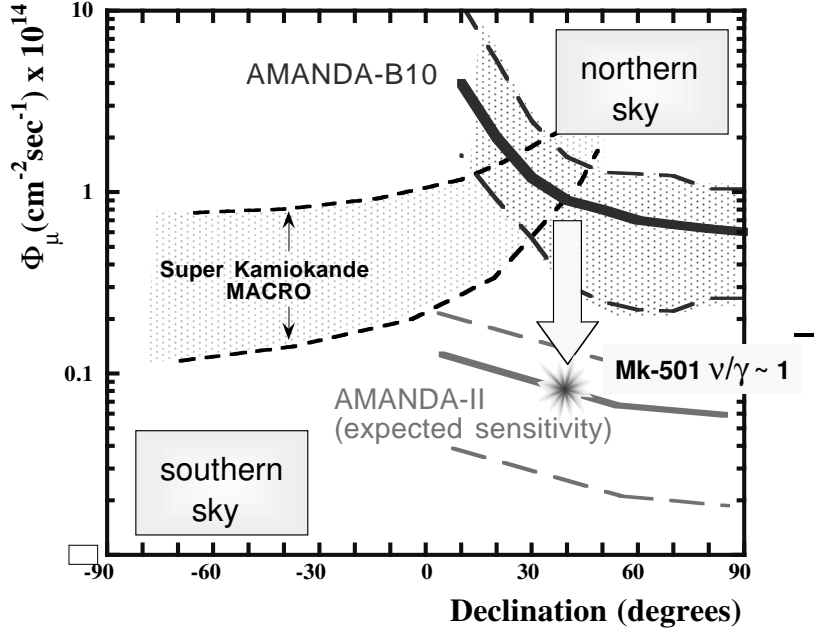}
\caption{Comparison of current flux limits from AMANDA-B10, SuperKamiokande\cite{sk}, and MACRO\cite{icrc01:macro}, and projected flux sensitivity for 600 live-days of AMANDA-II.  For reference, we indicate the neutrino flux from Markarian 501 assuming the neutrino flux and energy spectrum is identical to the observed high-state gamma ray flux.}
\label{fig:globflux}
\end{figure}

Figure~\ref{fig:globflux} summarizes the published experimental muon flux limits as a function of declination.  Bands indicate bin-to-bin variation at a given declination due to statistical fluctuation of background events.
As seen from Fig.~\ref{fig:globflux}, the sensitivity provided by AMANDA-II is sufficient to test the straightforward hypothesis that the brightest TeV gamma-ray sources emit neutrinos with the same flux and energy spectrum proportional to E$^{-2}$. One consequence of this hypothesis is that the observed ratio of neutrinos to gamma-rays may be larger than unity due to photon absorption within the source, or absorption enroute, but cannot be significantly less than unity because the neutrino cross-section is negligible compared to the photon cross-section. We highlight Markarian 501 in Fig.~\ref{fig:globflux}, which shows the neutrino flux limit from B10 along with the projected sensitivity of AMANDA-II with data on tape. If current generation neutrino telescopes do not detect neutrinos from these sources, then the neutrino flux can be can be adjusted through model-specific proton and electron efficiency factors to make models compatible with observational data.  Since model assumptions span a large range, it is possible that neutrino telescopes with an order of magnitude improvement in sensitivity will observe a positive signal.
 
\begin{table*}[htb]
\caption{Preliminary limits for selected point sources from 2000 data and estimated sensitivities (in  parenthesis) of AMANDA-II for data now available on tape ($\sim$600 live-days).  The sensitivity
         is defined as the predicted average limit from an
         ensemble of experiments with no signal, and is calculated using
         background levels predicted from off-source data. No systematic uncertainties are included.}
\label{table:point_source_sensitivities}
\renewcommand{\tabcolsep}{2pc} 
\renewcommand{\arraystretch}{1.2} 
\begin{tabular}{llll} \hline
Source              & \multicolumn{1}{c}{Declination} 
                            & \multicolumn{1}{c}{$\mu$ ($\times 10^{-15}$cm$^{-2}$s$^{-1}$)} 
                                   & \multicolumn{1}{c}{$\nu$ ($\times 10^{-8}$cm$^{-2}$s$^{-1}$)}\\ \hline

SS433               & 5.0   & 6.6 (3.1) & 4.2 (2.0) \\
Crab                & 22.0  & 6.9 (1.5)  & 7.0 (1.6) \\
Markarian 421       & 38.2  & 2.4 (0.8)  & 2.6 (0.9) \\
Markarian 501       & 39.8  & 1.9 (0.8)  & 2.2 (0.9) \\
Cygnus X-3          & 41.5  & 2.8 (0.8)  & 3.3 (0.9) \\
Cass. A             & 58.8  & 2.2 (0.7)  & 3.2 (1.0) \\ \hline  
\end{tabular}\\[2pt]
\end{table*}

The point source limits and sensitivities are calculated assuming approximately continuous emission over time.  However, many sources exhibit strong temporal variability in the electromagnetic bands.  If neutrino emission is strongly correlated with X-ray or gamma-ray variability~\cite{Atoyan,mBlazar} and the variability is well measured by orbiting or terrestrial telescopes, then sensitivity for a given source should improve with decreasing duration of the flaring episode. The degree of improvement is still under study.

\section{ATMOSPHERIC NEUTRINOS WITH AMANDA-II}
\begin{figure}[htb]
\includegraphics[scale=0.65]{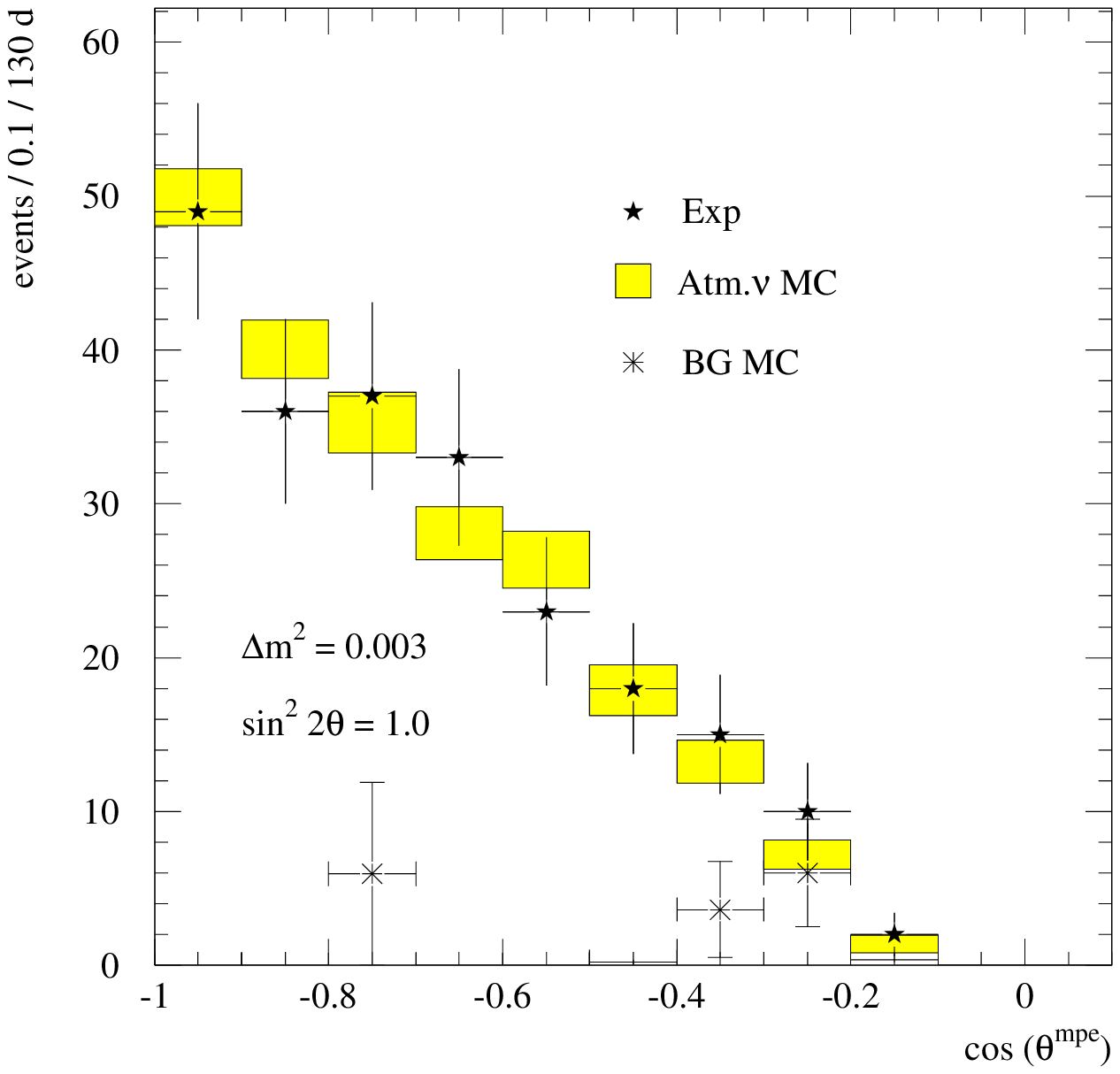}
\hfill  
\includegraphics[scale=0.43,bb=30 30 560 560]{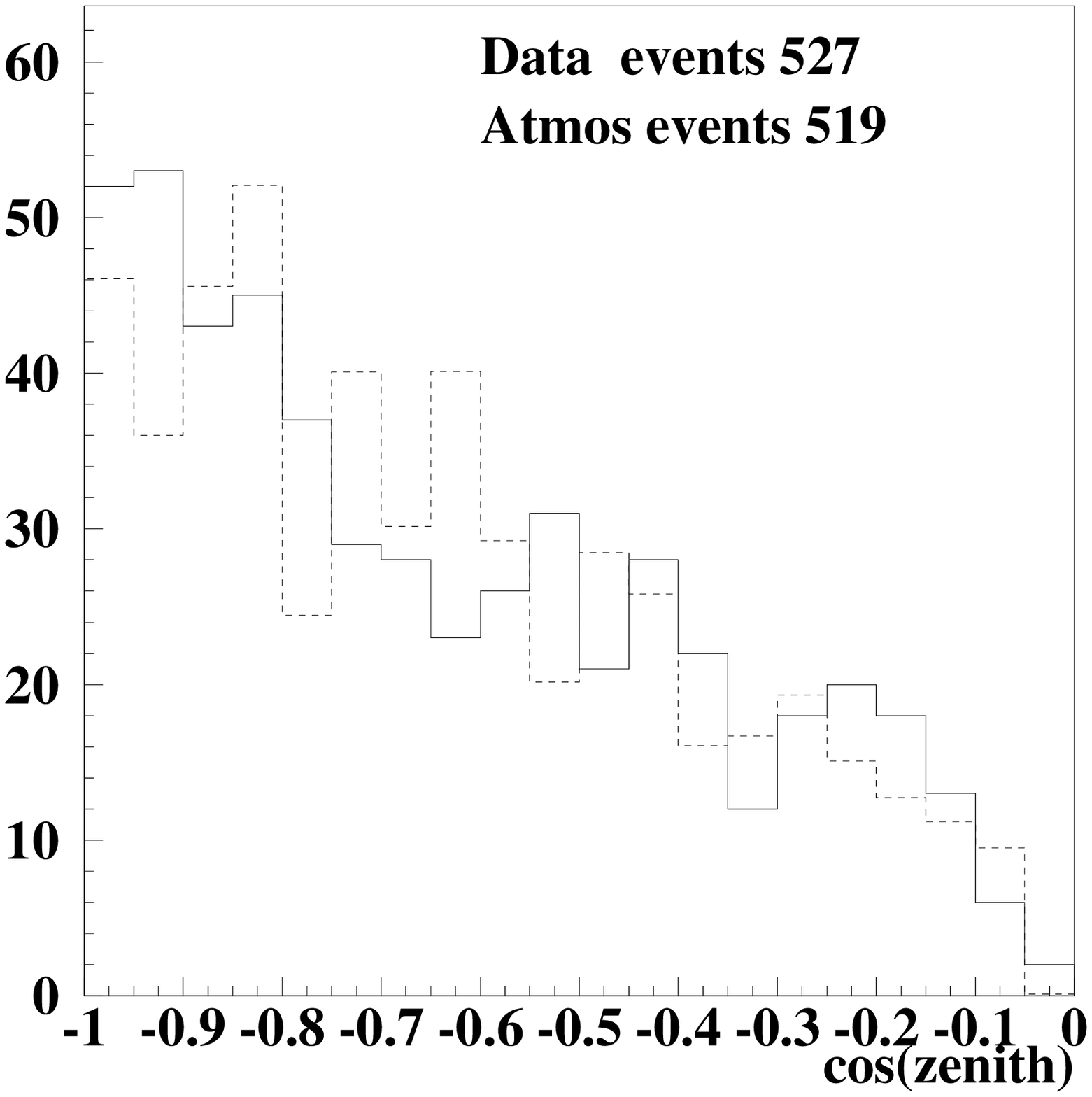}
\caption{{\bf Left:}Number of upward-going muon events in AMANDA-B10 data from 
         the year 1997, as a function of zenith angle ($\cos{\theta} =
         -1.0$ is vertically up in the detector).  The data are shown
         as dots and the Monte Carlo as boxes.  The simulation was performed
         with the neutrino oscillation parameters as indicated.
         The predicted signal efficiency is
         roughly 4\% and the background level is roughly 10\%, with both
         numbers improving near the vertical and degrading near the horizon.
         Simulations indicate that these events have an energy range
         given roughly by 60~GeV $< E_\nu <$ 300~GeV.}
\label{fig:b10-atmnu}

\caption{{\bf Right:}Number of upward-going muon events in AMANDA-II data from the 
         year 2000 as a function of zenith angle, using a preliminary
         set of selection criteria.  There are a total of 527 events
         in the data (solid line), and 519 events predicted by the
         atmospheric neutrino Monte Carlo (dashed line).  Simulations
         indicate that these events have an energy range given roughly
         by 100~GeV $< E_\nu <$ 1~TeV.  With more sophisticated selection criteria
         we expect improved response near the horizon.}
\label{fig:aii-atmnu}
\end{figure}

Without a known astronomical source of high energy neutrinos, detector calibration and consistency checks of the predicted response must be carried out by alternative methods. \textit{Atmospheric neutrinos}, being the only guaranteed source of high energy signal, provide an important tool to assess neutrino sensitivity.
A fraction of the atmospheric muon-neutrinos produced in the northern hemisphere by collisions with cosmic rays travel through the
earth, interact with the underlying earth or the ice near AMANDA, and
produce a muon which can be detected and reconstructed.  Using data
collected by AMANDA-B10 in 1997, we reconstructed roughly 300
upward-going muons which, as shown in Fig.~\ref{fig:b10-atmnu},
are in agreement with the predicted angular distribution.  This analysis focuses on obtaining a pure sample of low energy neutrinos to simplify comparison with simulation.  Since this analysis typically produces inefficiencies near the horizon, it is inappropriate to use for sources with harder energy spectra (typically, power law spectra proportional to $E^{-2}$ is targeted).  

A preliminary analysis of atmospheric neutrino data taken with
AMANDA-II demonstrates a substantial increase in the capability of the
enlarged detector.  Compared to the analysis using AMANDA-B10 data,
fewer selection criteria are required to extract a larger and
qualitatively cleaner sample of atmospheric neutrino-induced muons.
Figure~\ref{fig:aii-atmnu} shows the excellent shape agreement between
data and simulation achieved with a preliminary set of selection
criteria applied.  With more sophisticated selection criteria we
expect to see roughly twice the number of events shown in the figure
(corresponding to 2-3 times more events in AMANDA-II relative to
AMANDA-B10 for equivalent live-times) and we also anticipate improved
angular response close to the horizon.
%
%

\section{SEARCH FOR $\nu_\mu$ FROM DIFFUSE SOURCES WITH AMANDA-II}

The search for diffuse sources of UHE $\nu_\mu$--induced muons is
similar to the analysis used to detect atmospheric $\nu_\mu$--induced
muons, as both analyses require a sample of events with low contamination from
misreconstructed downward-going atmospheric muons.
Since high-energy muons will deposit more energy in the
detector volume than low-energy muons, the diffuse analysis further
requires that events have a high channel density, $\rho_{\rm ch} > 3$,
where the channel density is defined as the number of hit channels per
10~m tracklength.  The background in the signal region is estimated
by extrapolating from lower-energy data satisfying $\rho_{\rm ch} < 3$.

\begin{figure}[htb]
\includegraphics*[scale=0.42,bb=30 30 560 560]{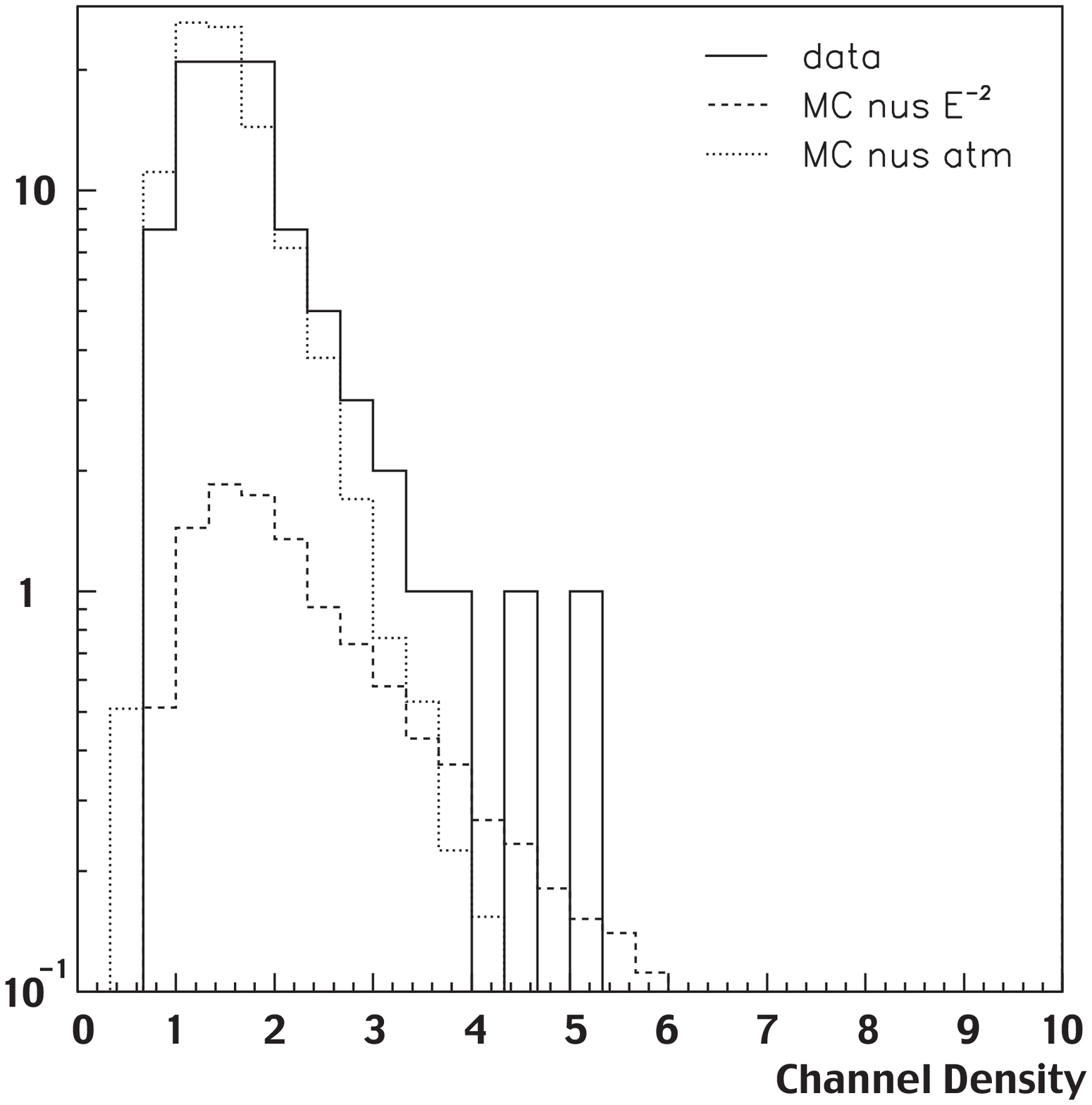}
\includegraphics*[scale=0.45]{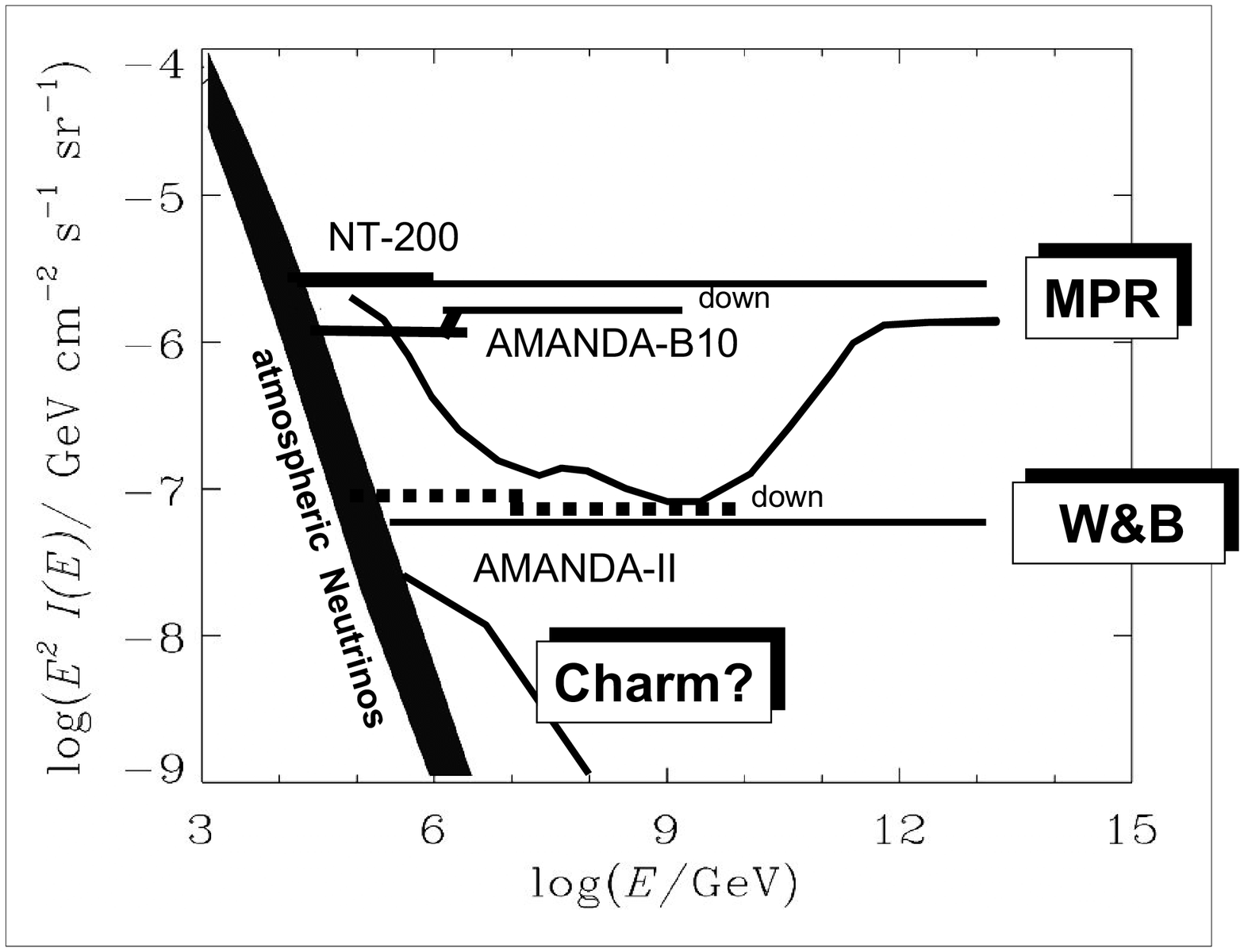}
\caption{{\bf Left:} Distributions of the channel density $\rho_{\rm ch}$ for data, 
         simulated signal and simulated background.  The simulated signal
         assumes a customary $E^{-2}$ power law spectrum at the source
         and a neutrino flux of
         $10^{-6}\,\mathrm{GeV\,cm^{-2}\,s^{-1}\,sr^{-1}}$.  Events are
         kept if they satisfy $\rho_{\rm ch} > 3$.}
\label{fig:aii-diffuse}

\caption{{\bf Right:} Experimental limits\cite{b10-diffuse}, projected sensitivities, and theoretical upper bounds\cite{MPR,WBbound} on the flux of high energy neutrinos from diffusely distributed sources. With the exception of NT-200 at Lake Baikal\cite{venice01:baikal}, all limits are derived for muon neutrinos. The preliminary lower energy AMANDA limits are derived from upward-traveling muons and the extreme high energy (EHE) limits are described by Hundertmark\cite{b10-ehe}. The horizontal length of the integral limits indicate the energy interval that contains $\sim$90\% of the events, assuming an energy spectrum proportional to $E^{-2}$. The broken curve of high energy background (labeled "Charm?") extending from the atmospheric neutrino curve estimates the additional contribution from prompt muons and neutrinos due to charm decay. }
\label{fig:diffuselim}
\end{figure}

Using a 20\% subsample of the AMANDA-II data from 2000, we detect 6
events satisfying all selection criteria.  Simulations indicate that
we would detect 3.0 events from a UHE neutrino flux at the current
best limit~\cite{b10-diffuse}, assuming a customary $E^{-2}$ power law
spectrum at the source, and 1.9 events from atmospheric neutrino
interactions.  (N.B. We use a subsample of the data in order to
achieve blindness in this analysis.)  The distributions of $\rho_{\rm
ch}$ for data, simulated signal and simulated background are shown in
Fig.~\ref{fig:aii-diffuse}.

The predicted average limit from an ensemble of experiments with no
signal, or {\it sensitivity}, is roughly $1.3 \times
10^{-6}$~GeV$\,$cm$^{-2}$s$^{-1}$sr$^{-1}$, and the preliminary limit is less
than roughly $10^{-6}$~GeV$\,$cm$^{-2}$s$^{-1}$sr$^{-1}$.  This is about the
same as the limit obtained with the {\it full} sample of AMANDA-B10
data from 1997.

The discovery potential of AMANDA-II (and other current generation neutrino telescopes) is especially high for diffuse signals.  The projected sensitivity of AMANDA-II in Fig.~\ref{fig:diffuselim} indicates that much of the available parameter space is probed. The space is bounded at low energies by the diffuse atmospheric neutrino and muon backgrounds. The background floor on this figure depends by the rather uncertain prompt muon and neutrino flux from charm production in the atmosphere.  At higher energies, neutrino absorption by the earth implies that diffuse signals originate predominantly from slightly above the horizon.  The AMANDA telescope can survey the extremely high energy (EHE) region of the neutrino energy spectrum using techniques that extract bright horizontal events from the less energetic atmospheric background\cite{b10-ehe}.  AMANDA can provide critical input to more sensitive techniques based on radio~\cite{ANITA}, fluorescence~\cite{EUSO-OWL}, air shower arrays~\cite{Auger}, and possibly acoustic~\cite{Acoustic} signatures from the cascades initiated by such highly energetic interactions. At this early stage of diffuse EHE analysis, the projected sensitivity contains substantial (perhaps factor of 2) uncertainty but AMANDA-II is poised to probe model predictions that lie above the MPR bound\cite{MPR} and those in the vicinity of the W\&B bound\cite{WBbound}. The sensitivity may be better than shown once the analysis techniques incorporate the waveform information that will become available during the next few years.

\section{SEARCH FOR $\nu_\mu$ FROM GRBs WITH AMANDA-B10 and -II}

\begin{figure}[htb]
\includegraphics[scale=0.45]{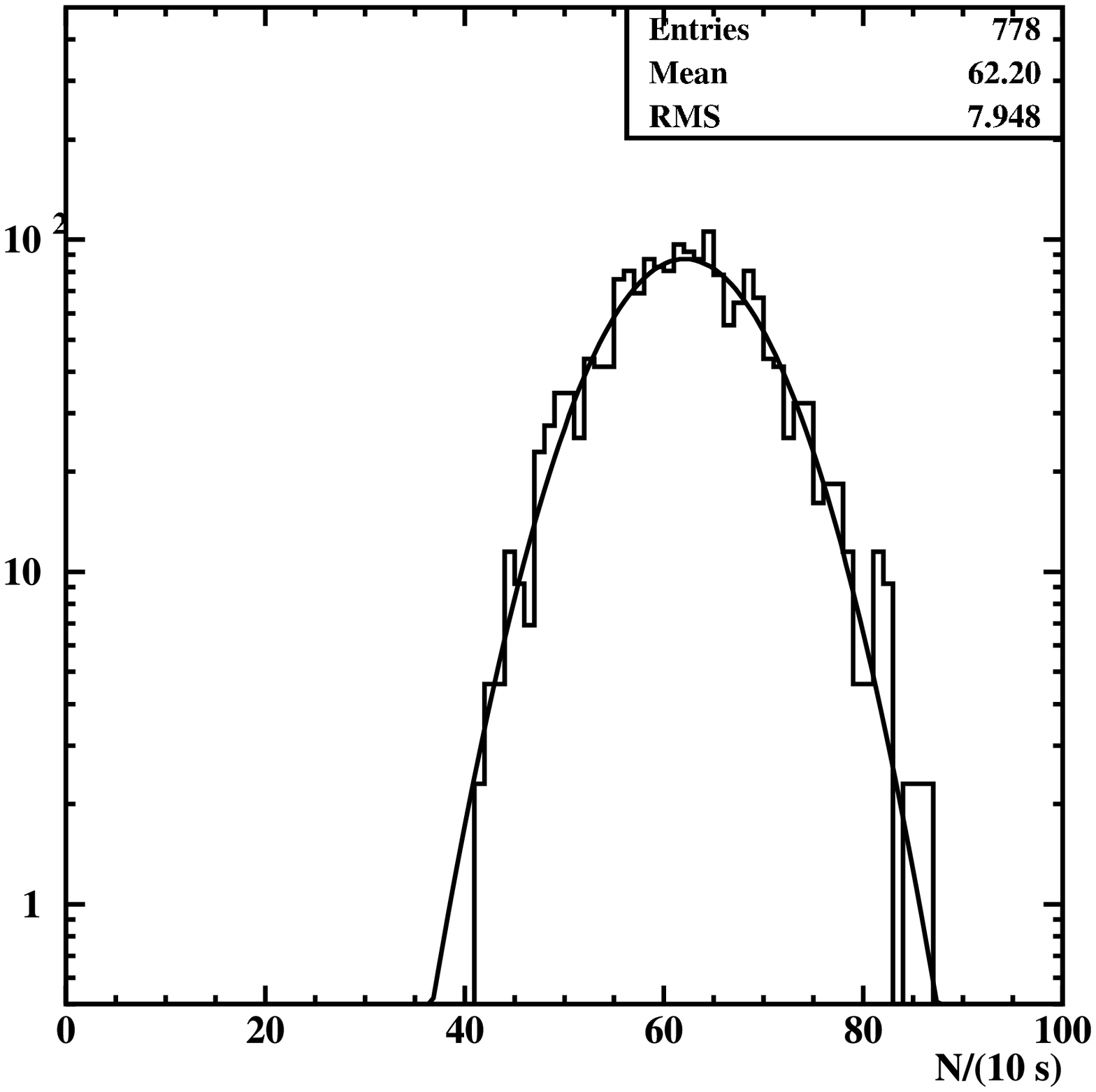}
\hfill
\includegraphics[scale=0.45]{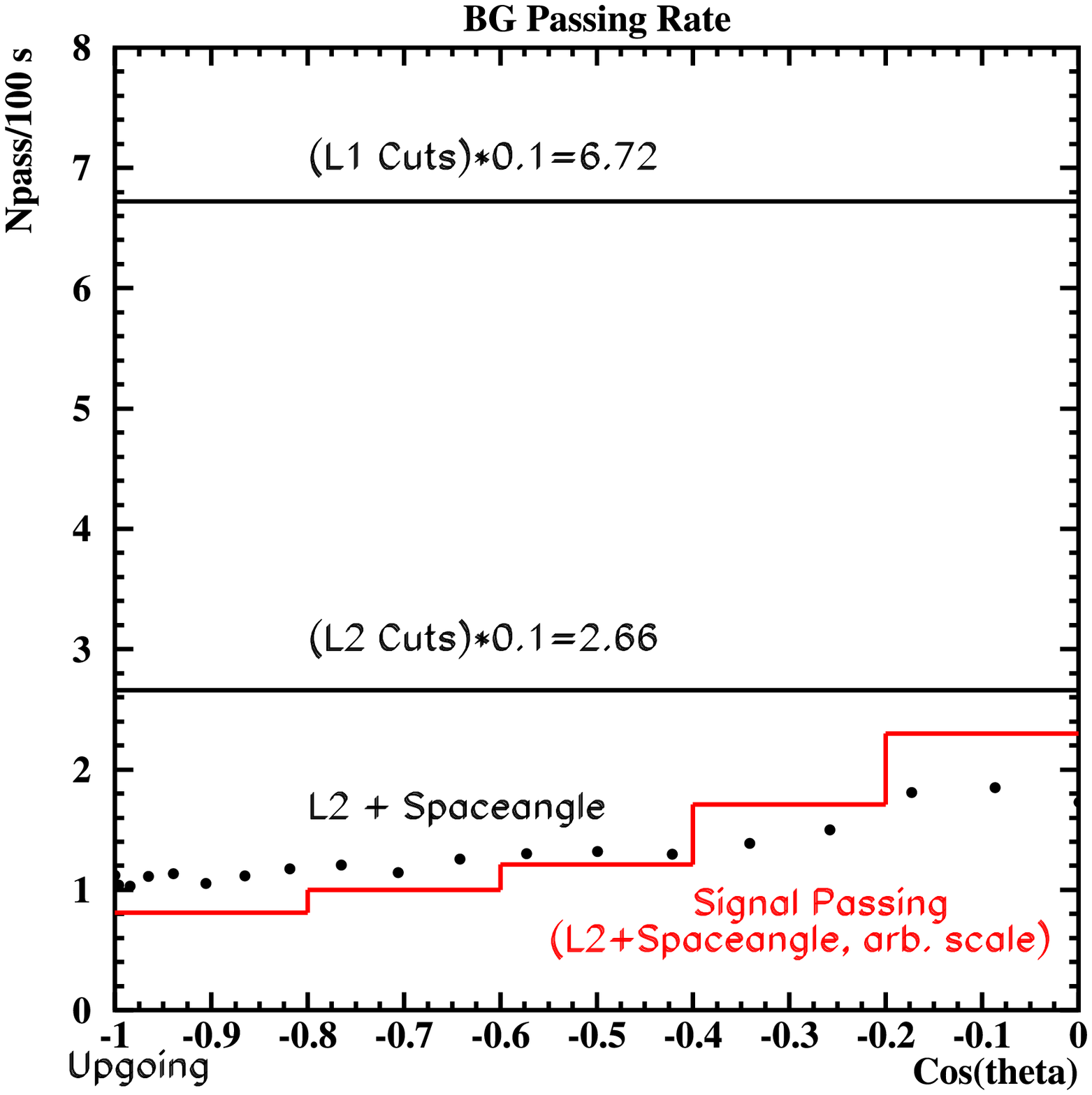}
\vspace{9pt}
\caption{{\bf Left:}The event count per 10~s period, demonstrating the good stability and gaussian fluctuation of the AMANDA-II off-time event rate.
         Some selection criteria have been applied.}
\label{fig:aii-stability}

\caption{{\bf Right:}Signal and background counts for the GRB per-burst analysis as a function of declination. Background counts for various stages of the analysis (L1 = level 1, L2= level 2, etc.) are determined for a time interval of 100 seconds, while the signal from the final stage of the analysis (L2+spaceangle)is arbitrarily scaled.}
\label{fig:GRB_BG}
\end{figure}
The search for UHE $\nu_\mu$--induced muons from gamma-ray bursts
(GRBs) utilizes temporal and directional information from
satellite-based observations of GRB photons to obtain very large effective area (the per-burst AMANDA-II analysis typically reaches 50\% of the maximum area determined by the hardware trigger conditions). Since AMANDA archives all of the data it collects, the search for correlated high energy neutrino emission uses the archival GRB database from BATSE.   The analysis reported here assumes that the neutrino emission occurs over the same time interval that contains 90\% of the gamma-ray photons in the GRB (T$_{90}$). We expand the window by a few seconds around T$_{90}$ to account for possible early emission as predicted by some models and to reduce the fine-tuning on the short burst population.  The relatively large detection area is a consequence of the modest background rejection, which requires approximately $10^{-4}$ for per-burst analysis and $\sim 2\times10^{-5}$ for composite searches.  This is contrasted with the approximately $10^{-6}$ rejection requirement for point source analysis and $\sim 10^{-8}$ for the diffuse source analysis.   Assuming the predicted
spectrum~\cite{grb-spectrum}, we search for muon neutrinos in the
energy range of 10--500~TeV and use off-source and off-time data to
estimate background and to achieve blindness in the analysis. 

Models of high energy neutrino emission from GRBs tend to focus on average properties, but several recent papers have asserted that rare, favorable fluctuations in the burst characteristics for one or two GRBs may dominate the neutrino flux.  This suggests a two-fold search strategy: one designed to maximize the individual (or per-burst) sensitivity and another designed to maximize the composite sensitivity for all measured GRB.  In this report, we focus on the simpler per-burst search strategy for the AMANDA-II analysis and note that  the results from the composite search using the 1997 data were reported elsewhere\cite{GRB97}.

The per-burst analysis of 2000 data from AMANDA-II looks for enhancements in the rate of upward-going muons over short periods from a fixed direction, with relatively loose background rejection requirements at the $10^{-4}$ level.
Event rate stability over short periods is therefore an important measure
of how effective this analysis can be.  Figure~\ref{fig:aii-stability}
shows the count rate per 10~s bin in a time window of roughly $\pm 1$~hour
around a particular GRB.  The agreement with a Gaussian
distribution shows that the detector did not experience instrumental
effects which could mimic a GRB.  Plots for all the other GRBs in the
sample exhibit the same well-understood behavior.  

\begin{figure}[]
\includegraphics[scale=0.4]{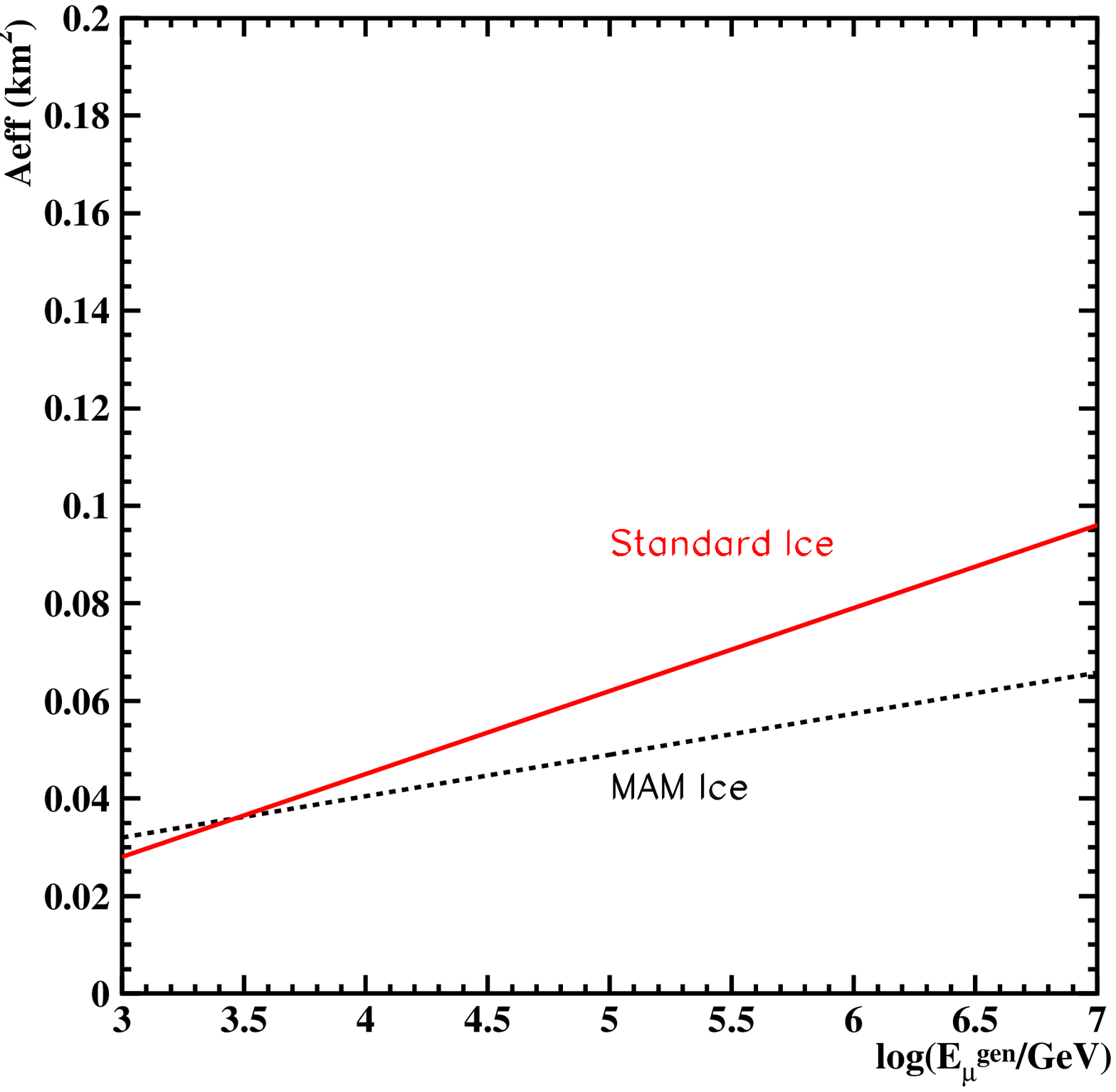}
\hfill
\includegraphics[scale=0.45]{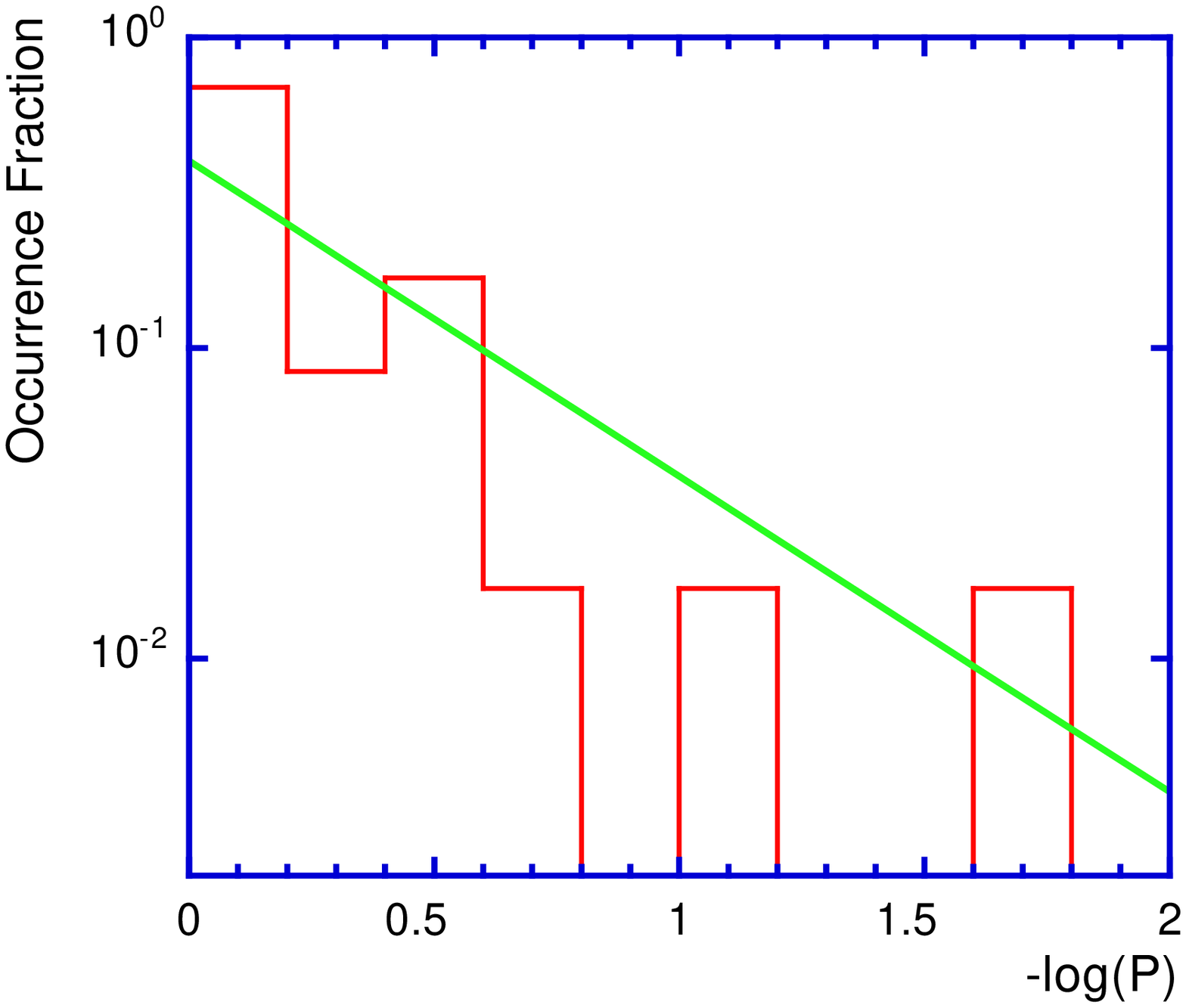}
\caption{{\bf Left:}The volume-averaged effective area as a function of the muon energy at production (typically a factor of 2 less than the incoming neutrino energy for charged current interactions).  The effective area for the GRB analysis shown for two ice models and also averaged over declination. Statistical uncertainty is about 15\% in this preliminary study.}
\label{fig:GRBarea}

\caption{{\bf Right:} Histogram of chance probabilities (P) to obtain experimentally observed counts during the T$_{90}$ window from a random fluctuation of expected background. The solid line was computed from off-time data of the GRBs by randomly redistributing the time stamps.  The time-shuffling procedure was repeated 100 times and the results averaged.}
\label{fig:GRBstat}
\end{figure}

All AMANDA-II data within the 2-hour window surrounding the GRB event were processed with full-iterative reconstruction of cleaned data. Cleaning procedures remove instrumental artifacts such as cross-talk signals. A very simple selection procedure achieved the requisite background rejection:  the reconstructed declination of the AMANDA-II event had to be larger than -10 degrees and the angular direction had to be within 22 degrees of the GRB direction obtained from BATSE. The angular dependence of the background events and detector sensitivity is shown in Fig.~\ref{fig:GRB_BG}.  At the final cut level, we include the effect of neutrino absorption by the earth for the assumed spectrum with a spectral break at 500 TeV. Since the angular dependence for both signal and background is similar, the analysis does not benefit from the additional complication of angle dependent cuts.  A total of 58 BATSE bursts were observed between February and May of 2000 (the sample includes triggered and non-triggered bursts to increase statistics). As shown in Fig.~\ref{fig:GRBarea}, the per-burst analysis reaches an effective area of 50,000 m$^2$ for $E_{\nu}$ = 100 TeV, approximately ~50\% of the effective area at trigger level.  From Fig.~\ref{fig:GRBstat} we conclude that search provides no evidence for correlated emission of high energy neutrinos from any GRB burst in the BATSE data sample from 2000.  The histogram agrees with the line that was computed assuming random fluctuation of off-time background events.   

In the AMANDA data
spanning the years 1997--2000, we anticipate having a sample of
roughly 500 GRBs to search through.  The longer term prospects look promising as well.  New satellites have either started operation in 2002 or planned for launch during the next half decade.  They will enhance the GRB-finding capabilities of the GRB Coordinates Network\cite{GCN}.  Within the same time frame, the GRB sensitivity of AMANDA-II is planned to be augmented by the first strings of IceCube\cite{albrechts-contribution}.  As few as 16 additional strings can achieve an effective area of 0.5 km$^2$ for neutrino energy in the theoretically interesting region around 100 TeV, although the potential of the configuration cannot match the full IceCube array. Figure~\ref{fig:PlusGRB} shows the volume-averaged effective area as a function of the muon energy at production.  In this simulation, the strings are separated by 125m and roughly centered on AMANDA-II. We point out that the configuration should achieve comparable detection area for generic searches for transient or episodic emission.

\section{SEARCH FOR DIFFUSE FLUX USING CASCADES IN AMANDA-B10 and -II}

In addition to methods based on elongated tracks, we have performed a full-reconstruction search for the Cherenkov light
patterns resulting from electromagnetic or hadronic {\it
cascades} induced by a diffuse flux of high-energy extraterrestrial
neutrinos~\cite{picrc,kowalski:diplm,taboada:phd,amanda-cascades}. 
Demonstrating $\nu$-induced cascade sensitivity is an important step
for neutrino astronomy because the cascade channel probes all neutrino
flavors, whereas the muon channel is primarily sensitive to $\nu_\mu$.
Compared to muons, cascades provide more accurate energy measurement
and better separation from background, but they suffer from far worse
angular resolution and reduced effective volume. It is more
straightforward to calibrate the cascade response of AMANDA through use of, for example, {\it in-situ} light sources.  As with muons, cascades become increasingly easier to identify and reconstruct as detector volumes get larger.

The electron neutrino produces cascades via the charged current
interaction and all neutrino flavors produce cascades via the neutral
current interaction. Cascade-like events are also produced in charged
current $\nu_\tau$ interactions.  The successful
reconstruction of pulsed laser data and the reconstruction of isolated
catastrophic muon energy losses, described in~\cite{amanda-cascades},
demonstrate that the detector is sensitive to high energy cascades.

\begin{figure}
\includegraphics[scale=0.42]{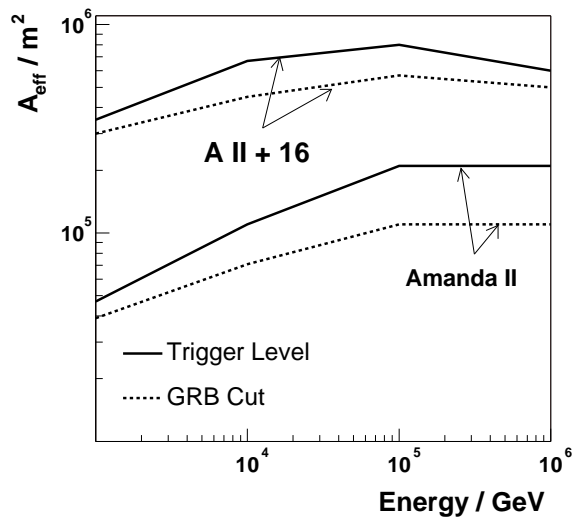}
\hfill
\includegraphics[width=0.45\textwidth]{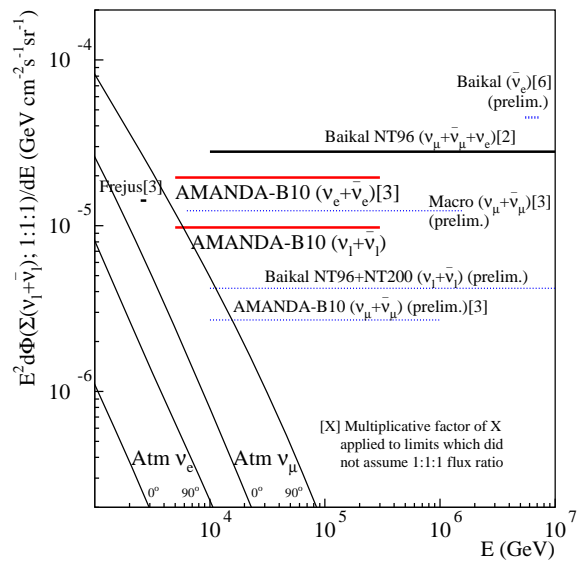}

\caption{{\bf Left:}The volume-averaged effective area as a function of the muon energy at production for AMANDA-II and 16 additional strings.  The effective area is averaged over declination. Solid lines correspond to hardware trigger with no background rejection. Dashed lines include background rejection to $10^{-5}$. Note that the trigger conditions for AMANDA-II in this study do not match the trigger conditions of the detector in 2000.}
\label{fig:PlusGRB}

\caption{\label{fig:limits} 
{\bf Right:}The limits on the cascade-producing neutrino flux, summed over the
three active flavors, presented for AMANDA and other experiments,
with multiplicative factors applied as indicated to permit comparison
of limits derived with different assumed neutrino fluxes at the
detector: Baikal ($\overline{\nu}_e$)~\cite{venice01:baikal} (at the
$W^\pm$ resonance); Baikal NT96
($\nu_\mu+\overline{\nu}_\mu+\nu_e$)~\cite{baikal:astropart}; Frejus
($\nu_\mu+\overline{\nu}_\mu$)~\cite{frejus}; MACRO
($\nu_\mu+\overline{\nu}_\mu$)~\cite{icrc01:macro}.  Baikal NT96+NT200
($\nu_l+\overline{\nu}_l$)~\cite{venice01:baikal,jan:baikal}; AMANDA B-10
($\nu_\mu+\overline{\nu}_\mu$)~\cite{b10-diffuse}; Also shown are the
predicted horizontal and vertical $\nu_e$ and $\nu_\mu$ atmospheric
fluxes~\cite{lipari}.}
\end{figure}

Our results together with other limits on the flux of diffuse
neutrinos are shown in Fig.~\ref{fig:limits}.  Since recent results
from other low energy neutrino
experiments~\cite{sno-cc,sno-nc,sno-dn,superk-atm} indicate that
high-energy extragalactic neutrinos will have a neutrino flavor flux
ratio of 1:1:1 upon detection, in this figure we scale limits derived
under different assumptions accordingly.  For example, compare a limit on the flux of
$\nu_e+\nu_\mu+\nu_\tau+\overline{\nu}_e+\overline{\nu}_\mu+\overline{\nu}_\tau$,
derived under the assumption of a ratio of 1:1:1, to a limit on just
the flux of $\nu_\mu+\overline{\nu}_\mu$, the latter must be degraded
by a factor of three.  (N.B.: We assume that
$\nu$:$\overline{\nu}$::1:1, and we take into account the different cross sections for $\nu$ and $\overline{\nu}$.)

Data acquired by AMANDA-II is currently under study and, as with the
analysis of atmospheric neutrinos and GRBs, preliminary results from that work
clearly demonstrate the enhanced power of the larger AMANDA-II
detector.  Angular acceptance improves to nearly $4\pi$, backgrounds
are much easier to reject, and energy acceptance improves by a factor
of three to $E_\nu \sim 1$~PeV. 

\section{LOOKING AHEAD}
	As the analysis from AMANDA-II data in year 2000 is refined for publication, the focus will start to shift to data acquired during 2001 and 2002, with a commensurate factor of three increase in statistical precision. We are also currently analyzing the data from AMANDA-B10 acquired in 1998 and 1999. 
	During the next two polar campaigns in Antarctica, the data acquisition system of AMANDA-II will be upgraded to handle much higher event rates and record the full signal waveforms from the optical modules. The added information will help to improve the energy response and angular resolution, especially for neutrinos with energies above PeV. Then, as the analysis programs are modified to incorporate the new detector capabilities, AMANDA-II will reach its full potential.  Beyond that, as few as 16 additional IceCube strings would dramatically boost the sensitivity of the system in several key physics goals and insure the continuation of state-of-the-art science output.   
\acknowledgments     
      
This research was supported by the following agencies: U.S.  National
Science Foundation, Office of Polar Programs; U.S. National Science
Foundation, Physics Division; University of Wisconsin Alumni Research
Foundation; U.S. Department of Energy; Swedish Research Council;
Swedish Polar Research Secretariat; Knut and Alice Wallenberg
Foundation, Sweden; German Ministry for Education and Research; U.S.
National Energy Research Scientific Computing Center (supported by the
Office of Energy Research of the U.S.  Department of Energy); FNRS-FWO, Flanders Institute to encourage scientific and technological research in industry (IWT); Belgian Federal Office for Scientific, Technical and Cultural Affairs (OSTC), Belgium; UC-Irvine AENEAS Supercomputer Facility; Deutsche
Forschungsgemeinschaft (DFG). D.F.C. acknowledges the support of the
NSF CAREER program. 


\end{document}